\documentclass[%showpacs,
showkeys,12pt,
preprint,preprintnumbers,nofootinbib,
groupedaddress,superscriptaddress,amsmath,amssymb]{revtex4}
\usepackage{graphicx}% Include figure files
\usepackage{dcolumn}% Align table columns on decimal point
\usepackage{bm}% bold math
\usepackage{amssymb}
\usepackage{amsmath}
\usepackage{epsfig}    
\usepackage{color}
\usepackage{slashed}
\usepackage{hhline}
\def\be{\begin{equation}}
\def\ee{\end{equation}}
\newcommand{\bea}{\begin{eqnarray}}
\newcommand{\eea}{\end{eqnarray}}
\newcommand{\nn}{\nonumber}

\numberwithin{equation}{section}

\begin{document}

%%%%%%%%%
%\title{Two loop neutrino model linking to leptoquark and diquark \\ with gauged hidden $U(1)$ symmetry}
\title{An Extended Colored Zee-Babu Model }
%\preprint{KIAS-P14078}
%

\author{Takaaki Nomura}
\email{nomura@kias.re.kr}
\affiliation{School of Physics, KIAS, Seoul 130-722, Korea}

%\author{Masaya Kohda}
%\email{mkohda@hep1.phys.ntu.edu.tw}
%\affiliation{Department of Physics, Chung-Yuan Christian University, Chung-Li 32023, Taiwan}

\author{Hiroshi Okada}
\email{macokada3hiroshi@gmail.com}
\affiliation{Physics Division, National Center for Theoretical Sciences, Hsinchu, Taiwan 300}

\date{\today}

\begin{abstract}
We study the extended colored Zee-Babu model introducing a vector-like quark and singlet scalar.
The active neutrino mass matrix and muon anomalous magnetic moment are analyzed, which can be fitted to experimental data satisfying the constraints from flavor changing neutral current.
Then we discuss signature of our model via vector-like quark production. In addition, the diphoton excess can be explained with the contribution from vector-like quark.
\end{abstract}
\maketitle
\newpage

 \section{Introduction}
 %%%
Radiative seesaw models are one of the interesting possibilities not only to generate active neutrino masses but also to explain some phenomenological viewpoints such as muon anomalous magnetic moment ($(g-2)_\mu$) and dark matter candidate~\footnote{In this paper, dark matter candidates are not included in.}, which are not still uncovered yet. Furthermore, these particles can be correlated with each other.
Thus a vast literature has recently arisen along this idea~\cite{Zee, Cheng-Li, zee-babu, Krauss:2002px, Ma:2006km, Aoki:2008av, Gustafsson:2012vj, Hambye:2006zn, Gu:2007ug, Sahu:2008aw, Gu:2008zf, Babu:2002uu, AristizabalSierra:2006ri, AristizabalSierra:2006gb,
Nebot:2007bc, Bouchand:2012dx, Kajiyama:2013sza,McDonald:2013hsa, Ma:2014cfa, Schmidt:2014zoa, Herrero-Garcia:2014hfa,
Ahriche:2014xra,Long1, Long2, Aoki:2010ib, Kanemura:2011vm, Lindner:2011it,
Kanemura:2011jj, Aoki:2011he, Kanemura:2011mw, Schmidt:2012yg, Kanemura:2012rj, Farzan:2012sa, Kumericki:2012bf, Kumericki:2012bh, Ma:2012if, Gil:2012ya, Okada:2012np, Hehn:2012kz, Baek:2012ub, Dev:2012sg, Kajiyama:2012xg, Kohda:2012sr, Aoki:2013gzs, Kajiyama:2013zla, Kajiyama:2013rla, Kanemura:2013qva,Law:2013saa, Dasgupta:2013cwa, Baek:2013fsa, Baek:2014qwa, Okada:2014vla, Ahriche:2014cda, Ahriche:2014oda,Chen:2014ska,
Kanemura:2014rpa, Okada:2014oda, Fraser:2014yha, Okada:2014qsa, Hatanaka:2014tba, Baek:2015mna, Jin:2015cla,
Culjak:2015qja, Okada:2015nga, Geng:2015sza, Okada:2015bxa, Geng:2015coa, Ahriche:2015wha, Restrepo:2015ura, Kashiwase:2015pra, Nishiwaki:2015iqa, Wang:2015saa, Okada:2015hia, Ahriche:2015loa, Ahn:2012cg, Ma:2012ez, Kajiyama:2013lja, Hernandez:2013dta, Ma:2014eka, Aoki:2014cja, Ma:2014yka, Ma:2015pma, Ma:2013mga,
%%%
radlepton1, radlepton2, Okada:2014nsa, Brdar:2013iea, Okada:2015nca, 
%%%
Okada:2015kkj, Fraser:2015mhb, Fraser:2015zed, Adhikari:2015woo, Kanemura:2015cca, Bonnet:2012kz,Sierra:2014rxa, Davoudiasl:2014pya, Lindner:2014oea,Okada:2014nea, MarchRussell:2009aq, King:2014uha, Mambrini:2015sia, Boucenna:2014zba, Ahriche:2016acx, Okada:2015vwh,
Kanemura:2015bli, Nomura:2016fzs, Nomura:2016seu, Okada:2016rav, Nomura:2016rjf, Ko:2016sxg, 
Ahriche:2016rgf, Nomura:2016run, Nomura:2016vxr, Lu:2016ucn, Kownacki:2016hpm, Thuc:2016qva, Ahriche:2016cio, Ahriche:2016ixu, Ma:2016nnn, Nomura:2016jnl, Cherigui:2016tbm, Hagedorn:2016dze, Chulia:2016ngi, Antipin:2016awv, Ma:2016zod, Balakrishna:1987qd, Balakrishna:1988bn,Arbelaez:2016mhg,Ma:1989tz,He:1989er, Nomura:2016emz}.
In many cases, neutrino mass is generated via loop diagram associated with colorless particles.
However colored particles also can propagate inside a loop diagram in neutrino mass generation, which would provide other phenomenologically interesting effects.

The excess of events in diphoton channel is announced by both ATLAS and CMS Collaborations where the invariant mass of diphoton is $m_{\gamma \gamma} \simeq 750$ GeV~\cite{Aaboud:2016tru, Khachatryan:2016hje}.
The production cross sections at $\sqrt{s}=13$ TeV are then indicated to explain the excess for narrow width case such that:
\begin{eqnarray}
\sigma(pp\to R\to \gamma\gamma) = \left\{ \begin{array}{c} 
                      5.5 \pm 1.5 {\text{ fb}~~\text{ATLAS~\cite{ATLAS:2016, Aaboud:2016tru}}}\,, \\
                      4.8 \pm 2.1 {\text{ fb}~~\text{CMS~\cite{CMS:2016owr,Khachatryan:2016hje}\,, \ \ }} 
                                            \end{array}\right. \label{eq:data_di}
\end{eqnarray}
where $R$ stands for the diphoton resonance. 
The best fit value of the width of $R$ is $\sim 45$ GeV by the ATLAS while narrow width is preferred by the CMS.
A candidate of $R$ is spin-0 or 2 particle and we consider a scalar particle $\phi$ with 750 GeV mass in this paper.  
The earlier works to interpret the excess can be referred to e.g.~\cite{Harigaya:2015ezk,Backovic:2015fnp,Angelescu:2015uiz,Nakai:2015ptz,Buttazzo:2015txu,DiChiara:2015vdm,Knapen:2015dap,Pilaftsis:2015ycr,Franceschini:2015kwy,Ellis:2015oso,Gupta:2015zzs,Kobakhidze:2015ldh,Falkowski:2015swt,Benbrik:2015fyz,Wang:2015kuj,Dev:2015isx,Allanach:2015ixl,Wang:2015omi,Chiang:2015tqz,Huang:2015svl,Ko:2016wce,Nomura:2016seu,Kanemura:2015bli,Cheung:2015cug,Nomura:2016fzs,Ko:2016lai}.

In this paper we extend the colored Zee-Babu model proposed in Ref.~\cite{Kohda:2012sr} by including isosinglet vector-like quark and SM singlet scalar field to explain $(g-2)_\mu$ and the diphoton excess.
The active neutrino matrix and $(g-2)_\mu$ are induced at two-loop and one-loop level.
We then analyze them taking into account the constraints from flavor changing neutral current (FCNC).
In addition, implications to collider physics are discussed where we focus on signature of newly introduced vector-like quark and explanation of the diphoton excess.

This paper is organized as follows.
In Sec.~II, we show our model,  including neutrino sector, $(g-2)_\mu$ and constrains from FCNC.
In Sec.~III, we discuss some implications of our model to collider physics including explanation of the diphoton excess.
We conclude and discuss in Sec.~IV.

%\newpage

%%%%%%%%%%%%%%%%%%%%%%%%%%%%%%%%%%%%%
%\section{The Model}
%\subsection{Model setup}
 \begin{widetext}
\begin{center} 
\begin{table}%[tbc]
%\begin{tiny}
\begin{tabular}{|c||c|c|c|c||c|c|}\hline\hline  
&\multicolumn{4}{c|}{Quarks} & \multicolumn{2}{c|}{Leptons} \\\hline
& ~$Q_{L_i}$~ & ~$u_{R_i}$~ & ~$d_{R_i}$ ~ & ~$Q'$ 
& ~$L_{L_i}$~ & ~$e_{R_i}$ 
\\\hline 
%%%
$SU(3)_C$ & $\bm{3}$  & $\bm{3}$  & $\bm{3}$& $\bm{3}$  & $\bm{1}$& $\bm{1}$   \\\hline 
%%%
$SU(2)_L$ & $\bm{2}$  & $\bm{1}$  & $\bm{1}$& $\bm{1}$  & $\bm{2}$& $\bm{1}$   \\\hline 
 %%%
$U(1)_Y$ & $\frac16$ & $\frac23$  & $-\frac{1}{3}$  & $-\frac{4}{3}$ & $-\frac12$  & $-1$\\\hline
 %%%
% $U(1)_H$ & $0$  & $0$ & $0$  & $x$ & $x$ & $2x$   \\\hline
%%%%
%$Z_2$ & $+$ & $+$  & $+$ & $-$ & $-$ & $-$  & $+$ & $+$ & $-$ & $-$ \\\hline
%%%
\end{tabular}
\caption{Field contents of fermions
and their charge assignments under $SU(3)_C\times SU(2)_L\times U(1)_Y$, where the lower index $i(=1-3)$ represents the number of flavors. }
\label{tab:1}
% \end{tiny}
\end{table}
\end{center}
\end{widetext}
\begin{table}[thbp]
\centering {\fontsize{10}{12}
\begin{tabular}{|c||c|c|c|c|}\hline\hline
%&\multicolumn{2}{c||}{Scalars (vevs$\neq 0$)} & \multicolumn{2}{c||}{Inert scalars}  & \multicolumn{2}{c|}{Colored scalars} \\\hline\hline 
  &~ $\Phi$ ~&~ $\varphi$ ~&~ $S^{a}_{\rm LQ}$  ~&~ $S^{ab}_{\rm DQ}$　%& $\varphi$ 
  \\\hhline{|=|=|=|=|=|}
%%%
$SU(3)_C$ & $\bm{1}$  & $\bm{1}$ & $\bm{3}$ & $\bm{6}$ \\\hline 
$SU(2)_L$ & $\bm{2}$   & $\bm{1}$  & $\bm{1}$ & $\bm{1}$ \\\hline 
$U(1)_Y$ & $\frac12$  & $0$ & $-\frac{1}{3}$ & $-\frac{2}{3}$   \\\hline
% $U(1)_H$ & $0$ & $-x$  & $-2x$  & $x$ & $4x$   \\\hline
%$Z_2$ & $+$   & $+$ & $-$ & $-$& $+$ \\\hline
\end{tabular}%
} 
\caption{Field contents of bosons
and their charge assignments under  $SU(3)_C\times SU(2)_L\times U(1)_Y$. }
\label{tab:2}
\end{table}

\section{ Model setup and analysis}
In this section, we devote to review our model.
Our field contents and their charge assignments are the same as
 the original colored Zee-Babu model proposed by Kohda,  Sugiyama, and Tsumura group in ref.~\cite{Kohda:2012sr}
 except for the vector-like quark $Q'$ with $SU(2)_L$ singlet and a gauge singlet boson $\varphi$.
 We show all the field contents and their charge assignments in Table~\ref{tab:1} for the fermion sector and  Table~\ref{tab:2} for the boson sector. Here $\varphi$ is expected to be a source of the 750 GeV boson for explaining the diphoton excess. 
 %recently reported by ALTLAS and CMS at the CERN Large Hadron Collider (LHC)~\cite{lhc}.
 The main motivation to introduce $Q'$ is to explain the sizable $(g-2)_\mu$, and obtain the sizable enhancement of the diphoton excess.~\footnote{we have checked that the original model cannot obtain enough diphoton excess as well as the sizable discrepancy of  $(g-2)_\mu$ from SM.} 
Under these framework, the renormalizably relevant Lagrangian is given by
\begin{align}
-{\cal L}^{}&=
(y_\ell)_{ij}\bar L_L \Phi e_{R_j} + (y_L)_{ij}\bar L^c_{L_i}(i\sigma_2) Q_{L_j} S_{LQ}^*+ (y_R)_{ij}\bar e^c_{L_i} u_{R_j} S_{LQ}^*+ 
(y_S)_{ij}\bar d^c_{R_i} d_{R_j} S_{DQ}^* \nn\\
&+Y_{i} \bar e_{R_i} Q'_{L} S_{LQ}^* + m_{Q'} \bar Q'Q'+{\rm h.c.} ,
\end{align}
where $\sigma_2$ is the second component of the Pauli matrix, and the second line is the new terms.
{ Since potential associated with leptoquark and diquark is trivial, we abbreviate the explicit expression; see the original paper~\cite{Kohda:2012sr} in details.}
Here we show the potential for $\Phi$ and $\varphi$ which determines the VEVs of them:
 \begin{align}
\label{eq:potential}
V & \supset  \mu^2  \Phi^\dagger \Phi + \lambda (\Phi^\dagger \Phi)^2+ m_\varphi^2 \varphi^2 + \mu_\varphi \varphi^3  + \lambda_\varphi \varphi^4 + \mu_{\varphi \Phi} \varphi (\Phi^\dagger \Phi) + \lambda_{\varphi \Phi} \varphi^2  (\Phi^\dagger \Phi)  \nonumber \\
& + (\text{terms containing leptoquark and diquark}).
 \end{align} 
The Higgs doublet $\Phi$ and singlet $\varphi$ are written by
\begin{equation}
\Phi = \begin{pmatrix} G^+ \\ \frac{1}{\sqrt{2}} ( v+ \tilde h + iG^0) \end{pmatrix}, \quad \varphi = \frac{1}{\sqrt{2}} (v_\varphi + \phi)\,, 
\end{equation}
where $G^+$ and $G^0$ are the Goldstone bosons, $\tilde h$ is the SM-like Higgs field,  and $v (v_\varphi)$ is the VEVs of $\Phi$ ($\varphi$). Then, applying the minimal conditions $\partial V(v,v_\phi)/\partial v =0$ and $\partial V(v,v_\varphi)/\partial v_\varphi =0$, we obtain the stable VEVs such that:
 \begin{align}
 &\mu^2 + \lambda v^2 + \frac{\lambda_{\varphi \Phi}}{2}  v_\phi^2 + \frac{\mu_{\varphi \Phi}}{\sqrt{2}} v_\phi = 0 \,, \nonumber \\ 
 & m^2_\varphi v_\varphi + \frac{3 \mu_\varphi v^2_\varphi}{2\sqrt{2}} + \frac{\mu_{\varphi \Phi} v^2}{2\sqrt{2}} + \lambda_\varphi v^3_\varphi + \frac{\lambda_{\varphi \Phi} v_\varphi v^2}{2} =0\,.
 \end{align}
 In our analysis, we assume $v_\varphi << v, m_\varphi$ so that the VEVs are approximated to $v \simeq \sqrt{- \mu^2/\lambda}$ and $v_\varphi \simeq - \mu_{\varphi \Phi} v^2/(\sqrt{2} (2 m_\varphi^2 + \lambda_{\varphi \Phi} v^2 ))$.
 Taking $|\mu_{\varphi \Phi}| \sim O(1)$ GeV, we can suppress mixing between SM Higgs and the singlet scalar to be consistent with experimental data regarding SM Higgs~\cite{Aad:2015gba,CMS:2014ega}.
 In the following analysis, we ignore the mixing effect.

\subsection{Neutrino sector}
The neutrino mass matrix is induced at the two-loop level as can be seen in the original paper~\cite{Kohda:2012sr}, and its formula is given by 
\begin{align}
{\cal M}_{\nu_{ab}}&\approx\frac{24\mu}{(4\pi)^4 m_{LQ}^2} \mu (y_L^*)_{ai} m_{d_i} (y_S)_{ij} m_{d_j} (y_L^\dag)_{jb} F_1(r) ,\\
F_1(r)&=\int_0^1dx \int_0^{1-x} dy \frac{1}{x+y(-1+y+r^2)}\ln\left[\frac{x+r^2 y}{y-y^2}\right],
\end{align}
where $\mu$ comes from the term of $S_{LQ}^* S_{LQ}^* S_{DQ}$, $r\equiv(m_{DQ}/m_{LQ})^2$, and we have assumed that the down quark masses in the loop are negligible compared to the masses of lepto-quark and di-quark bosons. Notice here that $F_1(r)$ is solved 
only through the numerical way, even though an approximated analytical formula is known in the limit of $r<<1$ or $r>>1$~\cite{AristizabalSierra:2006gb}. Neutrino oscillation data is given by diagonalize ${\cal M}_{ab}$ as follows:
\begin{align}
{\cal M}^{diag}_\nu=V_{MNS}^T {\cal M}_{\nu} V_{MNS},
\end{align}
where $V_{MNS}$ is the Maki-Nakagawa-Sakata mixing matrix of the neutrino.
As experimental values, we adopt the best fit values with the global analysis in ref.~\cite{Forero:2014bxa};
\begin{eqnarray}
&&  s_{12}^2 = 0.323, \; 
 s_{23}^2 = 0.567, \;
 s_{13}^2 = 0.0234,   \;
\delta_{CP} =1.34 \pi,
\\
&& 
%  m_{\nu_2} ({\rm eV}) = 0.0087,  \; 
  \ |m_{\nu_3}^2- m_{\nu_2}^2| =2.48 \times10^{-3} \ {\rm eV}^2,  \; 
 % m_{\nu_3} ({\rm eV}) = 0.0502 .
  \ m_{\nu_2}^2- m_{\nu_1}^2 =7.60 \times10^{-5} \ {\rm eV}^2, \nn
  \label{eq:neut-exp}
  \end{eqnarray}
where we assume one of three neutrino masses is zero with normal ordering for simplicity in the numerical analysis below.
%Here we prepare the specific textures on $y_L$ and $y_S$ to evade some constraints as we will see in the next subsection.

\subsection{Flavor Changing Neutral Currents  Lepton Flavor Violations}

Before discussing the Flavor Changing Neutral Currents (FCNCs) and the lepton flavor violations (LFVs),
we specify the textures of $y_L$ and $y_S$ to evade some  FCNC processes. In this paper, we adopt the following textures:
\begin{align}
y_L= V_{MNS} \times y_L' \equiv V_{MNS}\times \left[\begin{array}{ccc} (y_L)_{11} &0 & 0 \\
 0 & (y_L)_{22} & (y_L)_{23} \\
0 & (y_L)_{32} & (y_L)_{33} \\
  \end{array}
\right],\quad
%%%
y_S=\left[\begin{array}{ccc} 0 &0 & 0 \\
 0 & (y_S)_{22} & (y_S)_{23} \\
0 & (y_S)_{23} & (y_S)_{33} \\
  \end{array}
\right],
\end{align}
where $y_S$ is a symmetric matrix.
Then the neutrino mass formula can be simplified as follows:
\begin{align}
{\cal M}^{diag}_\nu &=V_{MNS}^T {\cal M}_{\nu} V_{MNS}=y^{'*}_L \omega y_L^{'\dag},\label{eq:reduced-neut}\\
\omega&\equiv\frac{24\mu}{(4\pi)^4 m_{LQ}^2} m_{d} y_S m_{d} F_1(r).
\end{align}
From Eq.(\ref{eq:reduced-neut}), one finds that $(y_S)_{23}$ is identically zero.

Here we discuss the Flavor Changing Neutral Currents (FCNCs) and the lepton flavor violations (LFVs), where we focus on terms related to the neutrino masses, that is, $y_L$ and $y_S$.
Under these above textures, {\it e.g.}, 
%the constraint on $D^+\to\pi^+ \mu^+\mu^-$ can be evaded from $(y_L)_{21}=0$, 
the constraint on $K^0-\bar K^0$ mixing and $B^0_d-\bar B^0_d$ mixing can be evaded  from $(y_S)_{11}=0$. 
%The constraints on $\mu\to e\gamma$ and $\tau\to e \gamma$ are also zero because $(y_L y_L^\dag)_{21}$ and $(y_L y_L^\dag)_{31}$ are zero.

Thus the bound on LFVs comes from the $\mu-e$ conversion and $\ell_i\to\ell_j \gamma$ process, and
The bounds on FCNCs come from $K^+\to\pi^+\nu\bar\nu$ and $D^+\to\pi^+\mu^+\nu^-$ decays and the $Q-\bar Q$ mixing, where $Q=K^0, B_s^0$.
These are respectively constrained by the following combinations~\cite{Carpentier:2010ue},
\begin{align}
&\mu-e\ {\rm conversion}:\quad \left|\frac{(y_L)_{21} (y_L^\dag)_{11}} {4\sqrt2 G_F m_{LQ^2}}\right|\lesssim 8.5\times 10^{-7},
\label{eq:const1}\\
&\ell_i\to\ell_j\gamma\ {\rm}:\quad \left|\frac{3\alpha_{em}|(y_Ly_L^\dag)_{ij}|^2C_i}{256\pi G_F^2 m_{LQ}^4}\right|\lesssim 
(4.2\times 10^{-13},3.3\times 10^{-8} ,4.4\times 10^{-8}) \\
& \hspace{7cm}  {\rm for}\quad  (i,j)=[(\mu,e),\ (\tau,e),\ (\tau,\mu) ],
\label{eq:mue}
\nn\\
&K^+\to \pi^+\nu\bar \nu\ {\rm }:\quad \left|\sum_{i,j}^{1-3} \frac{(y_L)_{i2} (y_L^\dag)_{1j}} {4\sqrt2 G_F m_{LQ}^2}\right|\lesssim 9.4\times 10^{-6},\\
%%%
&D^+\to \pi^+\mu^+\mu^-\ {\rm }:\quad \left| \frac{(y_L)_{21} (y_L^\dag)_{22}} {4\sqrt2 G_F m_{LQ}^2}\right|\lesssim 6.1\times 10^{-3},\\
%%%
&B^0_s-\bar B^0_s\ {\rm mixing}:%\quad \left|\sum_{i,j}^{1-3} \frac{(y_S)_{i2} (y_S^\dag)_{3j}} {2 G_F m_{DQ}^2}\right|
{\left| \frac{(y_S)_{22} (y_S^\dag)_{33}} { G_F m_{DQ}^2}\right| }
\lesssim 3.3\times 10^{-10},\label{eq:const5}
\end{align}
where $C_i\approx(1,1/5)$ for $(\mu,\tau)$. To numerical analysis, we will impose these above constraints.

\subsection{Muon Anomalous Magnetic Moment $(g-2)_\mu$}
First of all, the original paper suggests that the term through $y_L$ cannot induce the sizable $(g-2)_\mu$, assuming $y_R$ to be zero.
% the formula of $(g-2)_\mu$  is proportional to the following relation 
However we find that $y_R$ cannot contribute to $(g-2)_\mu$ even the case of $y_R \neq 0$ because of an accidental electric charge cancellation in the limit of massless mediating up-type quarks:
\begin{align}
\Delta a_\mu(y_R)\propto 2 Q_{S_{LQ}} + Q_{u_k}=2\times \left(-\frac{1}{3}\right)+\frac{2}{3}=0,
\end{align}
%%%
Thus we rely on the new source as a new term $Y$, and its form is given by
%and its analysis can be done independently because it does not contribute to the neutrino masses.
\begin{align}
&\Delta a_\mu(T)=Y^\dag_{2} Y_{2} m_\mu^2 \left(G_1[m_{Q'},m_{LQ}]+ 4 G_1[m_{LQ},m_{Q'}]\right),\\
&G_1[m_1,m_2]=
\frac{1}{24 m_1^2},\quad {\rm for }\ m_1=m_2,\\
&G_1[m_1,m_2]=
\frac{2 m_1^6+3 m_1^4 m_2^2-6 m_1^2 m_2^4 + m_2^6+12 m_1^4 m_2^2 \ln\left(\frac{m_2}{m_1}\right)}{12(m_1^2-m_2^2)^4},\quad {\rm for }\ m_1\neq m_2,
\end{align}
%%%
Also we have to consider the lepton flavor processes through this term.
Here we impose the constraints of $\ell_i\to\ell_j\gamma$, and its branching ratio ($BR$) is given by
\begin{align}
&BR(\ell_i\to\ell_j\gamma)=\frac{48\pi^3\alpha_{em} C_{i} }{G_F^2}
Y^\dag_{i} Y_{j} \left|G_1[m_{Q'},m_{LQ}]+ 4 G_1[m_{LQ},m_{Q'}]\right|^2,
\end{align} 
where $\alpha_{em}$ is the fine structure constant, and $G_F$ is the Fermi constant.
%, and $C_{i}\approx(1,1/5)$ for $(\mu,\tau)$.
Experimental bound for $BR(\ell_i\to\ell_j\gamma)$ is the same as the one in Eq.~(\ref{eq:mue}).

\subsection{Result of numerical analysis}

Applying the formulas in previous subsections, we numerically search for parameter region which can explain neutrino mass and $(g-2)_\mu$ satisfying all the flavor constraints.
In our numerical analysis, we have found the following allowed regions to satisfy all the constraints in Eqs.~(\ref{eq:const1}-\ref{eq:const5}) and obtain the sizable $(g-2)_\mu$($\sim{\cal O}(10^{-9})$): 
\begin{align}
& m_{LQ} \in [1000\,, 3000\,] \; \text{GeV},\quad \{m_{DQ},\mu,m_{Q'} \} \in [500\,, 3000\,]\; \text{GeV}, \nn\\
%\; \text{GeV},\quad  \mu \in [500\,, 3000]\ \text{GeV}, \nn\\
&  y_L \in [-0.3\,, 0.3]\, , \quad  Y_2 \in [1\,, \sqrt{4\pi}]\, , \quad  Y_{i\neq2} \in [-0.1\,,0.1],
	\label{eq:range_scanning}
\end{align} 
where three components of $y_S(\le \sqrt{4\pi})$ can directly be solved by applying the experimental values of the neutrino oscillations data with the best fit values in Eq.~(\ref{eq:neut-exp}).

\section{Implications to collider physics}

In this section, we explore the implications of our model to collider physics.
In our analysis, we focus on the signature of $Q'$ production and the explanation of the diphoton excess.

\subsection{Collider searches of $Q'$}
%%%%%%%%%%%%%%%%%%%%%%%%%%%%%%%%%%%%%%%%%%%%%%%%%%%%%%%%%%%%%%%%%%
\begin{figure}[tb] 
\begin{center}
\includegraphics[width=70mm]{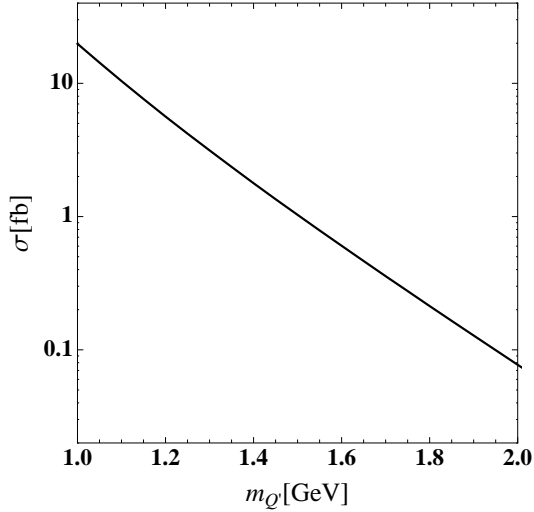} 
\caption{The cross section for $pp \to \bar Q' Q'$ at the LHC 13 TeV as a function of $m_{Q'}$.}
\label{fig:CX}
\end{center}
\end{figure}
%%%%%%%%%%%%%%%%%%%%%%%%%%%%%%%%%%%%%%%%%%%%%%%%%%%%%%%%%%
Here we discuss the production of $Q'$ at the collider, since 
$Q'$ has an interesting decay mode; $Q'\to (\ell)+S_{LQ}\to \ell+u_i$ through the Yukawa interactions associated with $Y_i$ and $y_{L(R)}$.
Since we have some freedom to chose $y_R$ while satisfying neutrino mass and flavor violating constraints, we here assume $BR$ for $S_{LQ} \to \ell u_i$ 
are universal for $\ell = e, \mu$ and $u_i = u, c$ while $BR$ for the other modes are negligibly small. 
On the other hand, the $Y_2$ is expected to be dominant to obtain sizable $(g-2)_\mu$ while avoiding the LFV constraints.
Thus $Q'$ dominantly decays into leptoquark and muon.
The $Q'$ pair is produced via QCD process.
We estimate the production cross section by use of CalcHEP~\cite{Belyaev:2012qa} implementing relevant interactions and applying CTEQ6L PDF~\cite{Nadolsky:2008zw} in the estimation. 
In Fig.~\ref{fig:CX} we show the production at the LHC 13 TeV as a function of $Q'$ mass $m_{Q'}$.

\begin{figure}[tb] 
\begin{center}
\includegraphics[width=70mm]{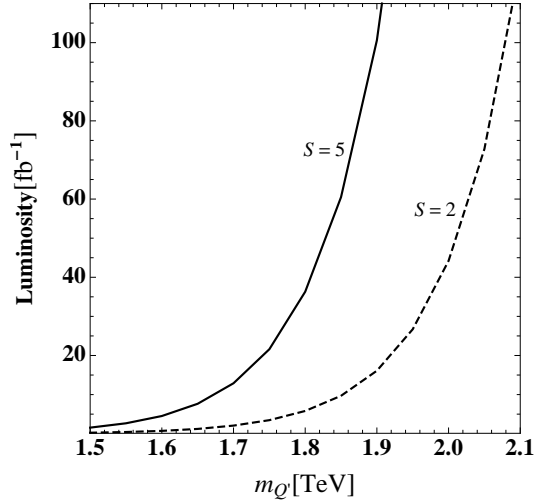} 
\caption{The luminosity required to obtain significance 5(2) shown as solid(dashed) line.}
\label{fig:lum}
\end{center}
\end{figure}

We then carry out simple simulation at the parton level with the event generator {\tt MADGRAPH/MADEVENT\,5}~\cite{Alwall:2014hca}, where the necessary Feynman rules and relevant parameters of the model are implemented by use of FeynRules 2.0 \cite{Alloul:2013bka} and the {\tt NNPDF23LO1} PDF~\cite{Deans:2013mha} is adopted.
As a signal, we generate the events for the process 
\begin{equation}
pp \to \bar Q' Q' \to S_{LQ} S_{LQ}^* \ell^+ \ell^- \to \ell^+ \ell^+ \ell^- \ell^- j j.
\end{equation}
Thus our signal includes four leptons and two jets; $4 \ell + 2 j$.
We then select the events as 4-leptons and jet(s) where number of jet is required to be $n_j \geq 1$.
For the SM backgrounds, we consider SM processes producing $ZZj$, $ZZjj$, $t \bar t Z$ and $Z Z W^\pm$ which can provide the 4-leptons plus jet(s) final states if $Z$ and $W^\pm$ decay into leptons and jets respectively.
In generating the events, the following basic cuts are adopted:
\begin{equation}
p_T (\ell) > 10 \, \text{GeV}, \quad p_T(j) > 20 \, \text{GeV}, \quad \eta(\ell) > 2.5, \quad \eta(j) > 5,
\end{equation}
where $p_T$ and $\eta$ are transverse momentum and pseudo rapidity respectively.
In addition, we apply selecting cuts of 
\begin{align}
\label{eq:cut2}
& p_T(\ell_{\rm leading}) > 50 \, \text{GeV}, \quad P_T(j_{\rm leading}) > 100 \, \text{GeV}, \nonumber \\
& m_Z - 10 \, \text{GeV} < M_{\ell^+ \ell^-} < m_Z + 10 \, \text{GeV} \quad (\text{veto}),
\end{align}
where $\ell_{\rm leading}(j_{\rm leading})$ are leading lepton(jet) and second line is for vetoing the region of $m_Z \pm 10$ for the invariant mass $M_{\ell^+ \ell^-}$. 
Then we estimate the significance for the signal by 
\begin{equation}
S = \frac{N_S}{\sqrt{N_B}}
\end{equation}
where $N_S$ and $N_B$ are the number of signal and background events respectively.
In table \ref{tab:events}, we show the number of events with basic cuts and after the cuts in Eq.~(\ref{eq:cut2}) for signal with $m_{Q'} = 1.5$ TeV and SM backgrounds using luminosity of 100 fb$^{-1}$.
We find that vetoing condition for $M_{\ell^+ \ell^-}$ highly suppress the SM backgrounds and the large significance can be achieved.
The luminosity required to obtain $S=5(2)$ is also shown in Fig.~\ref{fig:lum} as a function of $m_{Q'}$.
We see that the the significance of 5(2) can be reached for $Q'$ mass $m_{Q'} \lesssim 1.9(2.1)$ TeV could be tested at the LHC 13 TeV with the integrated luminosity of 100 fb$^{-1}$. 
In addition, the signature of $Q'$ can be seen as a bump in the distribution of invariant mass for two same sign lepton plus jets.
Note that taking into account detector efficiency the significance will be smaller, but our result is still reasonable estimation. 

 \begin{widetext}
\begin{center} 
\begin{table}%[tbc]
%\begin{tiny}
\begin{tabular}{|c ||  c | c c c c || c|}\hline
& signal ($m_{Q'} =1.5$ TeV) & $ZZj$ \qquad & $ZZ jj$ \qquad & $Z \bar t t$ \qquad &  $ Z Z W^\pm$ \qquad & $S$ \\ \hline
\# of signal events (basic cuts)  & 103. & 2.28$\times 10^3$  & 1.11$\times 10^3$  & 177.  & 7.98  & 1.72  \\ 
\# of signal events with Eq.~(\ref{eq:cut2}) & 93.2 & 0.84 &  0.96  & 3.54 & $< 0.1$  & 40.3  \\ \hline
%%%
\end{tabular}
\caption{The number of events for signal and backgrounds after kinematical cuts with the luminosity of 100 fb$^{-1}$ at the LHC 13 TeV. The significance is also shown in the last column.}
\label{tab:events}
% \end{tiny}
\end{table}
\end{center}
\end{widetext}
%
%%%%%%%%%%%%%%%%%%%%%%%%%%%%%%%%%%%%%%%%%%%%%%%%%%%%%%%%%%%%%%%%%%
%%%%%%%%%%%%%%%%%%%%%%%%%%%%%%%%%%%%%%%%%%%%%%%%%%%%%%%%%%

\subsection{Diphoton excess }
In this section we discuss the diphoton excess recently reported by ATLAS and CMS at LHC within our model.
Here $\varphi$ is identified as the source of 750 GeV, and diphoton processes are through leptoquark boson $S_{LQ}$, diquark boson $S_{DQ}$, and $Q'$ fermion mediating.   

Then we focus on the trilinear couplings of  $S_{LQ}(S_{DQ})$ and the Yukawa coupling of $Q'$ associated with $\phi$ which are given by
\begin{align}
\label{eq:phigg}
V_{} & \supset \lambda_{DQ \phi} \varphi^2 \left(S_{DQ}^\dagger S_{DQ} \right) + \lambda_{LQ \phi} \varphi^2 \left(S_{LQ}^\dagger S_{LQ} \right) 
+ \tilde \mu_{DQ \phi} \varphi \left(S_{DQ}^\dagger S_{DQ} \right) + \tilde \mu_{LQ \phi} \varphi \left(S_{LQ}^\dagger S_{LQ} \right) \nonumber \\
& \supset \lambda_{DQ \phi} v_\phi \phi \left(S_{DQ}^\dagger S_{DQ} \right) + \lambda_{LQ \phi}  v_\phi \phi \left(S_{LQ}^\dagger S_{LQ} \right) 
+ \frac{\tilde \mu_{DQ \phi}}{\sqrt{2}} \phi \left(S_{DQ}^\dagger S_{DQ} \right) + \frac{\tilde \mu_{LQ \phi}}{\sqrt{2}} \varphi \left(S_{LQ}^\dagger S_{LQ} \right) \nonumber \\
&\equiv \mu_{DQ} \phi \left(S_{DQ}^\dagger S_{DQ} \right) + \mu_{LQ} \phi \left(S_{LQ}^\dagger S_{LQ} \right), \\
L_{\rm Y} & = y_{Q'} \phi \bar Q' Q,
\end{align} 
where these interactions contribute to gluon fusion and diphoton decay processes of $\phi$ via $S_{LQ}$, $S_{DQ}$ and $Q'$ loops.
The loop induced effective coupling for $gg\phi$ can be written as 
\begin{equation}
\label{eq:Phigg}
{\cal L}_{gg\phi} = -\frac{\alpha_s}{8 \pi } \left( \frac{ 5 \mu_{DQ} }{2 m_{DQ}^2 } A_0(\tau_{DQ}) + \frac{ \mu_{LQ} }{2 m_{LQ}^2} A_0(\tau_{LQ}) + \frac{y}{2 m_{Q'}} A_{1/2} (\tau_{Q'}) \right) \phi G^{a \mu \nu} G^a_{\mu \nu}
\end{equation}
where $\tau_{X} = 4 m_{X}^2/m_\phi^2$ with $X = \{ LQ, DQ, Q' \}$ and factor $5$ in first term inside bracket comes from Casimir operator for {\bf 6} representation.
The loop functions are given by 
\begin{align}
 A_{1/2}(x) & = -2[x + (1-x) f(x))]\,, \nonumber  \\
 A_0(x)  & = x(1-x f(x))\,,
\end{align}
with $f(x) = \left[\sin^{-1} (1/\sqrt{x})\right]^2 $ for $x > 1$.

The diphoton decay $S \to \gamma \gamma $ is dominantly induced by charged leptoquark, diquark and vector-like quark loops where 
the partial decay width is given by
\begin{equation}
\Gamma_{\phi \to \gamma \gamma} \simeq \frac{\alpha^2 m_\phi^3}{256 \pi^3} \left|  \frac{8}{3} \frac{\mu_{DQ} }{2 m_{S_{DQ}}^2} A_0 (\tau_{DQ}) 
+  \frac{1}{3} \frac{\mu_{LQ} }{2 m_{LQ}^2} A_0 (\tau_{LQ}) +  \frac{16}{3} \frac{y_{Q'}}{ m_{Q'}^2} A_{1/2} (\tau_{Q'}) \right|^2,
\end{equation}
where color factor $N_c =6$ and $3$ are used for diquark and leptoquark respectively.
We also obtain the partial decay width for $S \to gg$ from effective interaction in Eq.~(\ref{eq:phigg});
\begin{equation}
\Gamma_{\phi \to gg} =  \frac{\alpha_s^2 m_\phi^2}{32 \pi^3} \left|  \frac{5 \mu_{DQ} }{ 2m_{DQ}^2 } A_0(\tau_{DQ}) + \frac{ \mu_{LQ}}{2 m_{LQ}^2} A_0(\tau_{LQ}) + \frac{y_{Q'}}{2 m_{Q'}^2} A_{1/2}(\tau_{Q'}) \right|^2.
\end{equation}
The total decay width is dominantly obtained from $\phi \to gg$ mode assuming $\lambda_{H\varphi} <<1 $ to suppress branching fraction for $\phi \to hh$.

\begin{figure}[tb] 
\begin{center}
\includegraphics[width=70mm]{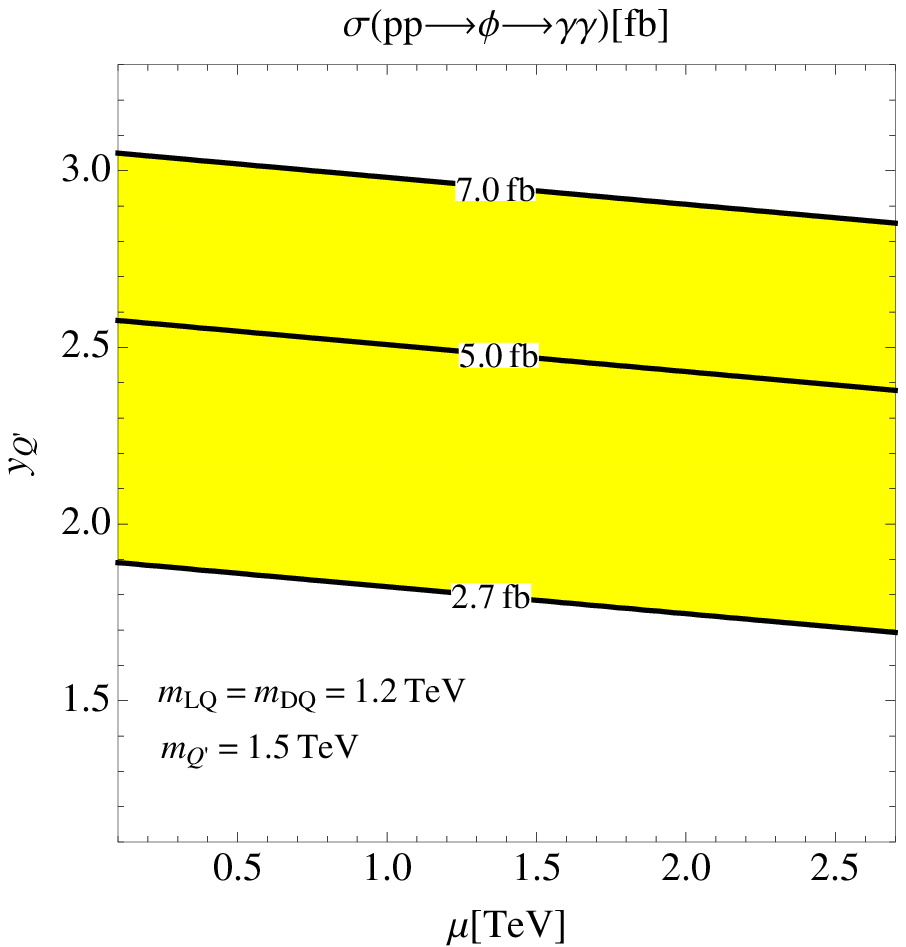} \hspace{5mm}
\includegraphics[width=70mm]{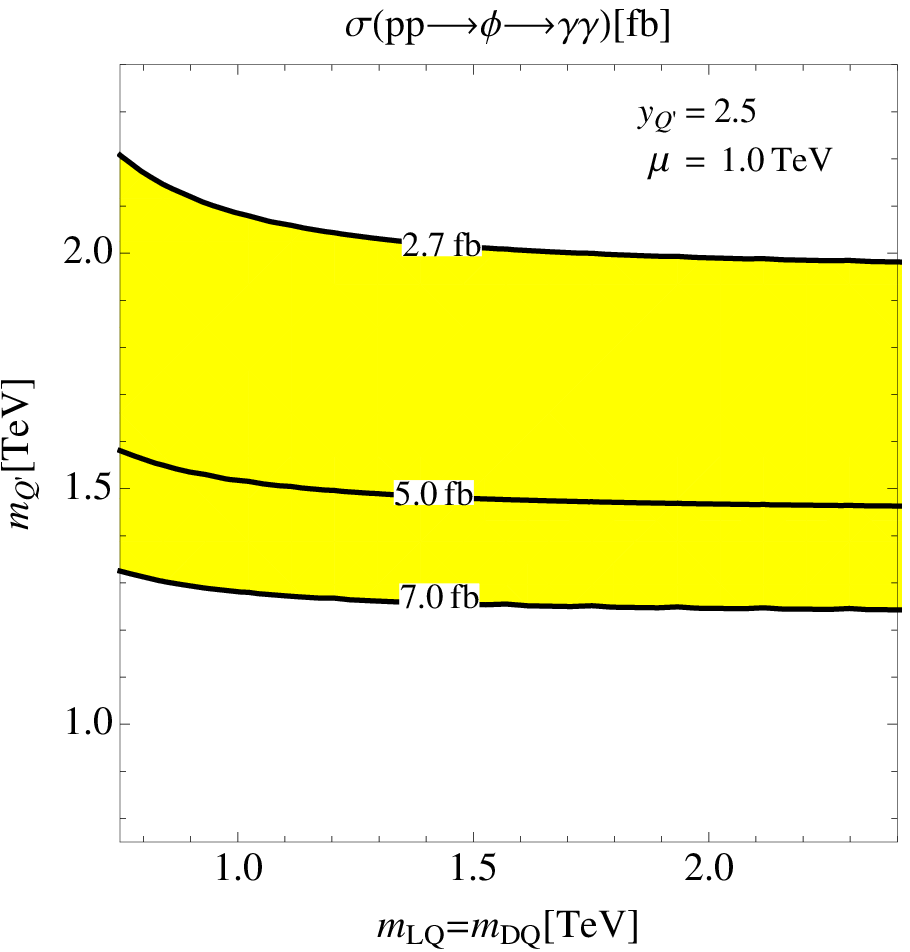} 
\caption{The cross section for $pp \to \phi \to \gamma \gamma$ which explain the diphoton excess in $\mu$-$y_{Q'}$ plane (left) and $m_{LQ(DQ)}$-$m_{Q'}$ plane (right) where the yellow colored regions satisfy Eq.~(\ref{eq:CXdiphoton}).}
\label{fig:diphoton}
\end{center}
\end{figure}

In the narrow width approximation, the cross section for the process $gg \to \phi \to \gamma \gamma$ can be expressed as~\cite{Franceschini:2015kwy} 
\begin{equation}
\sigma(gg \to \phi \to \gamma \gamma) \simeq \frac{C_{gg}}{s} \frac{\Gamma_{\phi \to gg}}{m_\phi} BR(\phi \to \gamma \gamma)
\end{equation}
where $C_{gg}$ is related to the gluon luminosity function, $s$ is the center of energy and $BR(\phi \to \gamma \gamma)$ is the branching fraction of $\phi \to \gamma \gamma$ decay.
For $\sqrt{s} = 13$ TeV, we adopt $C_{gg} \simeq 2137$.
In addition, we use the K-factor for gluon fusion production process as $K_{gg} \simeq 1.5$~\cite{Franceschini:2015kwy}. 
To explain the diphoton excess, the required cross section is 
\begin{equation}
\label{eq:CXdiphoton}
2.7 \, {\rm fb} \leq \sigma(gg \to \phi \to \gamma \gamma) \leq 7.0 \, {\rm fb},
\end{equation}
taking into account 1$\sigma$ error of ATLAS and CMS results in Eq.~(\ref{eq:data_di}).
In Fig.~\ref{fig:diphoton}, we show the parameter region which provide the cross section Eq.~(\ref{eq:CXdiphoton}) in $\mu$-$y_{Q'}$ and $m_{LQ(DQ)}$-$m_{Q'}$ planes 
where we take $\mu \equiv \mu_{LQ} = \mu_{DQ}$ and fixed some parameters as indicated in the plots.
We find that $O(1)$TeV trilinear coupling and $O(1)$ Yukawa coupling $y_{Q'}$ can explain the diphoton excess.
%%%%%%%%%%%%%%%%%%%%%%%%%%%%%%%%%%%%%%%%%%%%%%%%%%%%%%%%%%%%%%%%%%
%%%%%%%%%%%%%%%%%%%%%%%%%%%%%%%%%%%%%%%%%%%%%%%%%%%%%%%%%%

\section{ Conclusions and discussions}

In this paper we have discussed extended colored Zee-Babu model in which isosinglet vector-like quark and singlet scalar are included to accommodate $(g-2)_\mu$ and the 750 GeV diphoton excess indicated by ATLAS and CMS experiments. Then active neutrino mass matrix and $(g-2)_\mu$ are analyzed, which are induced at two loop and one loop level respectively. 
We also took into account the flavor changing neutral current to check the consistency with experimental constraints. 
Our analysis have shown that neutrino mass matrix and $(g-2)_\mu$ can be fitted with experimental data satisfying the constraints.

We also explored implications of our model to collider physics focusing on vector-like quark signature and the diphoton excess.
The vector-like quarks are produced via QCD process and decays into lepton and leptoquark which decays into lepton and jet.
Thus the signal of vector-like quark pair production is four-leptons plus jets.
We then estimated the production cross section and carried out simple simulation study including SM background.
By adopting some kinematical cuts, we have shown the vector-like quark which have mass $m_{Q'} \lesssim 2$ TeV can be tested at the LHC 13 TeV with integrated luminosity $100$fb$^{-1}$. 
The diphoton excess is explained by 750 GeV SM singlet scalar boson $\phi$ which couples to colored particles: diquark, leptoquark and vector-like quark.
We then find that the $\sigma (gg \to \phi \to \gamma \gamma) \sim 5$ fb can be obtained with perturbatively consistent couplings and $O(1)$ TeV colored particles.

%\section*{ Appendix}
%%%%%%%%%%%%%%%%%%%...

%\newpage
%%%%%%%%%%%%%%%%%%%%%%%%%%%%%%%%%%%
%\hspace{0.2cm} {\bf Acknowledgments}
%\section*{Acknowledgments}:
%\vspace{0.5cm}
\section*{Acknowledgments}
\vspace{0.5cm}
Authors would like thank Dr. Masaya Kohda for fruitful discussions.
H. O. is sincerely grateful for all the KIAS members, Korean cordial persons, foods, culture, weather, and all the other things.
%%%%%%%%%%%%%%%%%%%%%%%%%%%%%%%%%%%
%%%%%%%%%%%%%%%%%%%%%%%%%%%%%%%%%%%

\end{document}